\documentclass[prl,superscriptaddress,amsfonts,amssymb,
amsmath,showpacs,twocolumn]{revtex4}
\usepackage{epsfig}
\usepackage{graphics}
\date{\today}

\begin{document}

\title{Awaking the vacuum in relativistic stars}
\author{William C.\ C.\ Lima}\email{william@ursa.ifsc.usp.br}
\affiliation{Instituto de F\'\i sica de S\~ao Carlos,
Universidade de S\~ao Paulo, Caixa Postal 369, 15980-900, 
S\~ao Carlos, SP, Brazil}
\author{George E.\ A.\ Matsas}\email{matsas@ift.unesp.br}
\affiliation{Instituto de F\'\i sica Te\'orica, Universidade 
Estadual Paulista,
Rua Dr.\ Bento Teobaldo Ferraz, 271 - Bl.\ II, 01140-070, 
S\~ao Paulo, SP, Brazil}
\author{Daniel A.\ T.\ Vanzella}\email{vanzella@ifsc.usp.br}
\affiliation{Instituto de F\'\i sica de S\~ao Carlos,
Universidade de S\~ao Paulo, Caixa Postal 369, 15980-900, 
S\~ao Carlos, SP, Brazil}

\pacs{04.40.Dg, 04.62.+v, 95.30.Sf}

\begin{abstract}
Void of any inherent structure in classical physics, the {\it vacuum}
has revealed to be incredibly crowded with all sorts of processes in 
relativistic quantum physics.
Yet, its direct effects are usually so subtle that its 
structure remains almost as evasive as in classical physics. Here, in 
contrast, 
we report on the discovery of a {\it novel} effect
according to which the vacuum is {\it compelled} to 
play an unexpected central role in an astrophysical context. We show that 
the 
formation of relativistic stars may lead the vacuum energy density of a 
quantum field to an exponential growth. The
vacuum-driven evolution which would then follow may lead to unexpected 
implications for astrophysics, while the observation of stable 
neutron-star configurations 
may teach us much on the field content of our Universe.
\end{abstract}

\maketitle

Born from the efforts to formulate a consistent relativistic version 
of quantum mechanics, the successful formalism of {\it quantum field 
theory} (QFT) has unveiled an incredibly rich structure for what was 
thought in classical physics to be the most uninteresting of the states: 
the {\it vacuum}. From the early Dirac sea of negative-energy states to 
the picture of virtual particles constantly being created and annihilated, 
the vacuum has acquired conceptual importance for a consistent description 
of nature. However, from the observational point of view, its existence 
continues to be almost as evasive as in classical physics, demanding 
carefully designed experiments in order to detect its subtle effects. 
(The {\it Casimir effect}~\cite{Casimir48}, according to which two (or more) 
objects experience a force between them solely due to the fact that their 
presence changes the vacuum energy in a position-dependent way, is such 
an example, successfully observed in laboratory. See, e.g., 
Refs.~\cite{Chan01,Lamoreaux05}.)
Particularly interesting is the fact that the vacuum, being a dynamical 
entity, {\it gravitates}, which using the concepts of general relativity (GR)
means that it affects and is affected by the geometry of the spacetime.
This rests at the root of the {\it Hawking effect}~\cite{Hawking74,Hawking75}, 
according to 
which black holes should emit a thermal bath of particles. Although this
semiclassical-gravity effect of ``black hole evaporation''
has become a rare landmark for those who 
seek a complete quantum gravity theory,
it is virtually unobservable for realistic 
astrophysical black holes. Here, in contrast, 
we investigate possible astrophysical
implications of a  recently-proposed
semiclassical-gravity mechanism~\cite{LV} by which
the vacuum of {\it free} quantum fields is {\it forced}
to play a 
{\it central} role. We focus on the
context of {\it relativistic stars} and show that the formation of 
(realistic) compact objects may disturb the vacuum of a quantum field 
in a way which leads its energy density to an {\it exponential} growth. 
Eventually, this vacuum energy should 
take control over the evolution of the background spacetime, determining the 
ultimate fate of the relativistic star and possibly leading to unexpected 
implications for astrophysics. On the other hand,
the observation of stable neutron-star configurations may be used 
to rule out
the existence of fields for which the effect should have been 
triggered.


{\bf Vacuum energy in gravitational fields:} 
The task of calculating ``vacuum energies'' of quantum fields in gravitational 
backgrounds is one which is properly addressed (though not easily performed in 
general) {\it only} by using techniques of {\it quantum
field theory in curved spacetimes} (QFTCS), which is the minimal and most 
conservative way of merging the successful formalisms of QFT and 
GR~\cite{Parker68,Parker69,BD82,Fulling89,Wald94,Parker09}. 
Born more than four decades ago~\cite{Parker68,Parker69}, 
QFTCS has proven its value by providing the proper context in which legitimate 
``low-energy'' (i.e., below Planck scale) quantum gravitational phenomena,
like the black hole evaporation, could be unveiled.
Therefore, before presenting the arguments which lead to our main result, 
let us give the reader a brief overview on how to obtain the vacuum energy of a 
quantum field in a curved background using the canonical approach to QFTCS. 
In a nutshell, it goes as follows: (i) one solves the generally covariant field 
equation for a complete set of normal modes $\{u_\alpha^{(+)},u_\alpha^{(-)}\}$ 
of the field $\Phi$ 
in a {\it given} {\it fixed} 
background geometry with appropriate boundary/initial conditions which express 
the choice of the (Fock) Hilbert space of interest; (ii) one (formally) expands 
the 
field operator $\hat\Phi$ using these normal modes with the operator 
coefficients of the positive-norm $u_\alpha^{(+)}$ (negative-norm
$u_\alpha^{(-)}$) modes being interpreted as the 
annihilation (creation) operators for that particular mode; (iii) then one 
(formally)
constructs the stress-energy-tensor operator $\hat T_{\mu \nu}$ out of its 
classical expression by substituting $\Phi$ by  $\hat\Phi$, 
and calculates its expectation value in the state of interest (usually the 
vacuum of the chosen Fock Hilbert space). Since $\hat T_{\mu \nu}$ is a 
quadratic 
expression in  $\hat\Phi$ (which should be viewed as an operator-valued 
distribution), some regularization and renormalization procedure must be 
applied 
in step (iii) in order to give rise to an expectation value which is free 
from {\it ultraviolet} divergences.
 
Because the background spacetime has to be fixed already in step (i), 
the procedure outlined above is expected to give meaningful results whenever 
the backreaction of the expectation value of $\hat T_{\mu \nu}$ on the 
background spacetime (through the semiclassical Einstein equations) can be 
neglected. Notwithstanding, {\it failure} to comply with this condition can 
be used to expose situations where semiclassical-gravity effects become 
crucial. This is the strategy we follow here. 


{\bf Disturbing the vacuum during the formation of a star:}
For the sake of simplicity, we follow Ref.~\cite{LV} and 
consider a real scalar field $\Phi$ with 
mass $m\geq 0$ 
and arbitrary coupling $\xi$ with the Ricci scalar curvature $R$. 
The field equation is the usual Klein-Gordon equation with the additional 
scalar-curvature coupling:
$\left(-\square +m^2+\xi R \right)\Phi=0$.
(We adopt natural units in which $\hbar=c=1$ unless stated otherwise.) 
We consider the case in which matter initially scattered throughout space 
with very low density eventually collapses to form a static (and stable) 
spherically symmetric compact object (according to {\it classical} GR).
Therefore, the background metric can be taken to be asymptotically flat
in the past and asymptotically static and spherically symmetric in the future:
\begin{eqnarray}
ds^2\sim
\left\{
\begin{tabular}{ll}
$-dt^2+d\vec{x}^2$ & \!\!\!, past\\
$f (-dt^2+ d\chi^2)+r^2 (d\theta^2+\sin\theta^2 d\varphi^2)$ &  
\!\!\!, future
\end{tabular}
\right. 
\label{LE}
\end{eqnarray}
where $f=f(\chi)>0$ and $r=r(\chi)\geq 0$ are functions of the coordinate 
$\chi$ alone, with $f(\chi)\to 1$  and  $r(\chi)/\chi \to 1$ for 
$\chi\to \infty$, and
$dr/d\chi>0$ so that no trapped light-like surface is
present.

The natural in-vacuum state is the one 
associated with the normal modes $u^{(+)}_{\vec{k}}$
and $u^{(-)}_{\vec{k}}\equiv (u^{(+)}_{\vec{k}})^\ast$ which in the 
asymptotic past take the form of the usual flat-space stationary modes:
\begin{equation}
u^{(+)}_{\vec{k}}
\stackrel{{\rm past}}{\sim}
(16\pi^3 \omega_{\vec{k}})^{-1/2}\;
e^{-i(\omega_{\vec{k}} t-\vec{k}\cdot \vec{x})},
\label{inmode}
\end{equation}
where $\vec{k}\in {\mathbb R}^3$ and $\omega_{\vec{k}}:=\sqrt{\vec{k}^2+m^2}$ 
is 
interpreted as the energy of the mode.
Since the spacetime is also asymptotically static in the future, after the 
compact 
object is formed, there is a 
natural out-vacuum state, which is the one associated with a different set of 
normal modes $v^{(+)}_{\varpi l \mu}$ and 
$v^{(-)}_{\varpi l \mu}\equiv (v^{(+)}_{\varpi l \mu})^\ast$ 
which in the asymptotic future assume the stationary form
\begin{equation}
v^{(+)}_{\varpi l \mu}
\stackrel{{\rm future}}{\sim}
[\sqrt{2\varpi}\;f(\chi)r(\chi)]^{-1}
e^{-i\varpi t}F_{\varpi l}(\chi)Y_{l\mu}(\theta,\varphi),
\label{outmode}
\end{equation}
where  $Y_{l\mu}$ ($l=0,1,2,...$  and $\mu=-l,...,l$ ) are the usual 
spherical-harmonic 
functions and
$F_{\varpi l}$  are solutions of Schr\"odinger-type equations
\begin{equation}
\left(
-d^2/d\chi^2
+V^{(l)}_{\rm eff}
\right) F_{\varpi l}=\varpi^2 F_{\varpi l}
\label{Geq}
\end{equation}
satisfying usual boundary conditions and normalization. 
Here, $\varpi$ is the energy of the mode (which no longer needs to be 
greater than $m$) according to static observers in the asymptotic future which 
are far 
away from the compact object.
The effective potential  $V^{(l)}_{\rm eff}=V^{(l)}_{\rm eff}(\chi)$ 
is given by
\begin{equation}
V^{(l)}_{\rm eff}=f \left[
m^2+\xi R +l (l+1)/r^2
\right]+ 
(1/r) d^2 r/d\chi^2.
\label{Veff}
\end{equation}

The functions  $f(\chi)$ and $r(\chi)$ can be related to the matter/energy 
distribution of the compact object through the Einstein equations of GR. 
For instance, assuming the compact object to be made of perfect fluid 
(assumption which will be used throughout this article for simplicity), 
the effective potential can be put into the form
\begin{equation}
V^{(l)}_{\rm eff}=f \left[
m^2+(\xi-1/6)R +\frac{l(l+1)}{r^2}
+
\frac{8 \pi G}{3}\left( 
\bar{\rho}-\rho
\right)
\right],
\label{Veffrho}
\end{equation}
where  $\rho=\rho(\chi)$ is the matter/energy density of the compact object 
and $\bar{\rho}=\bar{\rho}(\chi)$ represents its {\it averaged} value 
up to $\chi$: $\bar{\rho}(\chi):=3 M(\chi)/(4 \pi r(\chi)^3)$,
with $M(\chi)$ being the mass of the object up to $\chi$. 
(Recall also that, according to GR, $R=8 \pi G (\rho-3 p)$, 
where $p$ is the pressure which sustains the configuration of the compact object 
and $G$ is Newton's gravitational constant.) 

In general the in- and out-vacuum do not coincide, which means that particles 
are 
created
due to the change in the gravitational background. However,
this particle creation is usually negligible.
In contrast, 
as recently pointed out in a quite general context~\cite{LV}, there is another 
possible influence of the gravitational background on the quantum vacuum: 
well-behaved
gravitational fields may trigger an exponential increase of the vacuum energy 
density
leading to a vacuum-dominated scenario.
Applied to the context of compact objects,
the gravitational fields which are entitled to ``awake'' the vacuum energy
are the ones for which the effective potential in Eq.~(\ref{Veffrho}) 
gets sufficiently {\it negative} somewhere in the asymptotic future so that 
Eq.~(\ref{Geq}) possesses well-behaved solutions $F_{\varpi l}$  for
$\varpi^2=-\Omega^2<0$. In this case, additional, {\it non-stationary}, 
normal modes $w^{(+)}_{\Omega l \mu}$ and $w^{(-)}_{\Omega l \mu}\equiv
(w^{(+)}_{\Omega l \mu})^\ast$  with the asymptotic behavior
\begin{equation}
w^{(+)}_{\Omega l \mu}
\stackrel{{\rm future}}{\sim}
\frac{\left(e^{-\Omega t + i \pi/12}+e^{\Omega t- i \pi/12}\right)}{\sqrt{2 
\Omega}}
\frac{F_{\varpi l}(\chi)}{r(\chi)}Y_{l\mu}(\theta,\varphi)
\label{voutmode}
\end{equation}
are {\it necessary} in order to expand the field in the asymptotic future. This 
implies
that at least some of the in-modes eventually go through a 
phase of exponential growth,
which leads to an exponential 
increase of the vacuum expectation value of $\hat \Phi^2$. This is then 
reflected 
in an exponential growth of the vacuum expectation value of $\hat T_{\mu \nu}$ 
(see Eqs.~(10--13) of Ref~\cite{LV}).
(It 
is important to note that this {\it asymptotic divergence} 
has nothing to do with high-frequency modes, so that renormalization plays 
no important role in this analysis.)
As a consequence, no matter what the value 
of the classical energy density of the compact object is, eventually the 
{\it vacuum} 
energy density (and pressure, and momentum density) of the field will 
dominate and 
take control over the spacetime evolution. 


{\bf Sharp transition to vacuum dominance
in relativistic stars:} 
It is interesting to note that in principle 
the vacuum-dominance effect
could be triggered for massive fields with $m^2 \ll R\sim G\rho$ (so that a
region where $V_{\rm eff}^{(l)}$ is negative might exist for
values of $|\xi|$ not too large). However,
recovering the units in which these quantities are usually expressed, this 
condition implies
\begin{equation}
\frac{m^2/(3.5 \times 10^{-12}\;{\rm eV})^2}{\rho/(10^{14}\;{\rm g}/{\rm cm}^3)}
\ll 1.
\label{mcondition}
\end{equation}
Therefore, considering that relativistic stars can reach densities 
$\rho \sim 10^{14-17}$~${\rm g}/{\rm cm}^3$~\cite{Lattimer07}, 
the most ``natural'' mass satisfying this is $m=0$, which we shall assume 
in what follows.
\begin{figure}[ht]
\centering
\epsfig{file=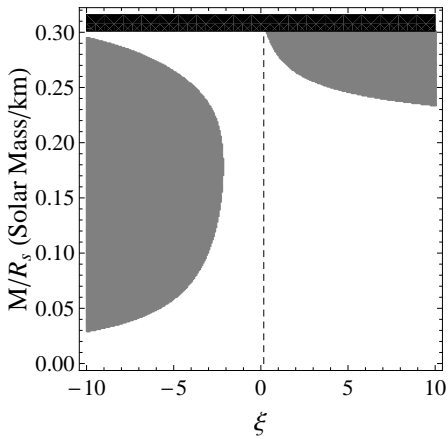,scale=0.7}
\caption{Diagram showing the values of the ratio $M/R_s$ of uniform-density
compact objects which trigger the exponential growth of 
the vacuum energy density of
massless scalar fields with coupling $\xi$ (dark-gray region).
The vertical dashed line indicates the conformal-coupling value $\xi=1/6$.
The black region represents the values of $M/R_s$ for which no equilibrium 
can be classically obtained.}
\label{Fig1PRL}
\end{figure}
\begin{figure}[ht]
\centering
\epsfig{file=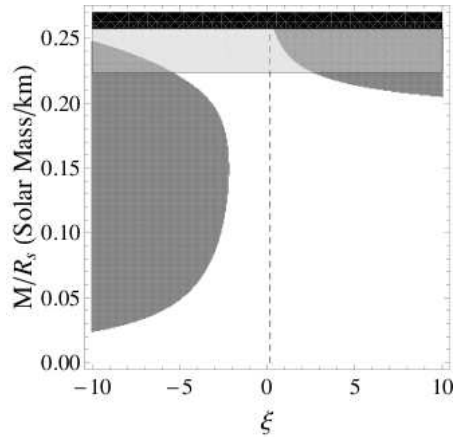,scale=0.7}
\caption{Same as diagram in Fig.~\ref{Fig1PRL} but now for
compact objects with parabolic
density profile.
The light-gray region represents 
values of
$M/R_s$ for which equilibrium configurations are classically unstable for most 
{\it realistic}
equations of state (see, e.g., Ref.~\cite{Lattimer01}).}
\label{Fig2PRL}
\end{figure}

It may come as a surprise to verify that the
simplest distribution possible for an idealized compact object, 
the {\it uniform} energy-density distribution, is already entitled to awake the
vacuum energy of quantum fields.
Assuming
$\rho=\bar{\rho}\equiv \rho_0$,  
the Tolman-Oppenheimer-Volkoff (TOV) equation of relativistic hydrostatic 
equilibrium
can be analytically integrated to obtain the pressure distribution $p$ and then 
the function
$f$ (see, e.g., Ref.~\cite{Wald84}). Then, substituting these into 
Eq.~(\ref{Veffrho}),
a numerical search for ``bound'' solutions of Eq.~(\ref{Geq}) with 
$\varpi^2 < 0$
(and $l=0$, which is the most promising case) can be performed for 
several values of the radius $R_s$ and the mass $M=4 \pi \rho_0 R_s^3/3$
of the compact object. In this uniform-density 
case, the existence of such ``bound'' solutions 
depends only on the ratio $M/R_s$ and the value of $\xi$. The results of this 
search 
are shown in 
Fig.~\ref{Fig1PRL}. We see that a massless field with any value of $\xi>1/6$ or 
$\xi \alt -2$ can have its vacuum energy density exponentially 
amplified for some range of 
$M/R_s$.

This idealized uniform-density 
case  serves to illustrate that well-behaved 
background geometries can indeed induce the vacuum-dominance effect
for fields with appropriate masses and scalar-curvature couplings. This, 
however, is not enough if we want to explore possible observational
implications. For that matter, more
realistic density and pressure profiles for the compact object have to be 
used. Although there is a certain amount of uncertainty on the correct 
equation of state for matter at densities as high as 
$10^{15-17}$~${\rm g}/{\rm cm}^3$, the density profile
of realistic neutron stars with masses $1.5 M_{\odot}\sim 2.2 M_{\odot}$
can be closely approximated by a parabolic radial dependence,
$\rho=\rho_c (1-r^2/R_s^2)$~\cite{Lattimer01}, where $\rho_c=15 
M/(8 \pi R_s^3)$
is the central density. In this case,
the TOV equation can also be 
analytically integrated to obtain the pressure distribution $p$ and the 
background metric~\cite{Tolman39}, and
a numerical search for ``bound'' solutions of Eq.~(\ref{Geq}) with 
$\varpi^2 < 0$
(and $l=0$)
can be performed, as in the previous case.
The existence of ``bound'' solutions again
depends only on the ratio $M/R_s$ and the value of $\xi$. 
The results of this search 
are shown in 
Fig.~\ref{Fig2PRL}. 
Note that Figs.~\ref{Fig1PRL}~and~\ref{Fig2PRL} are very similar but it 
is worthwhile to point out that in this more realistic case the effect is
triggered for smaller values of $M/R_s$. In particular, 
values of $M/R_s$ for which 
most realistic equations of state are causal~\cite{Lattimer01}, and
which represents classically stable configurations, can awake the 
vacuum of fields with proper couplings.

The time scale which governs the 
exponential growth of the vacuum energy density, energy flow and pressure 
is given by  $\Omega^{-1}$ (which usually is of order 
$|V_{\rm eff}^{(l)}|^{-1/2}$). 
In fact, the {\it dominant} contribution to the vacuum energy density in the 
asymptotic 
future can be estimated from 
Ref.~\cite{LV}:
\begin{eqnarray}
\rho_V
&\stackrel{{\rm future}}{\sim}&
\Omega h(\bar{r})e^{2\Omega t}/R_s^3
\nonumber \\
&\sim&\!\!\!\!
h(\bar{r})
\exp\left(
\frac{t/(10^{-5}~{\rm s})}{R_s/(10~{\rm km})}
\right)
\frac{ 10^{-62}~{\rm g}/{\rm cm}^3}{R_s^4/(10~{\rm km})^4},
\label{rhov}
\end{eqnarray}
where $h(\bar{r})$ is a dimensionless function of  
$\bar{r}:= r/R_s$ and we have used that for a bound solution
$\Omega\sim 1/R_s$. 
The exact values assumed by $h(\bar{r})$ 
depend on the details of the spacetime evolution between the
two static 
asymptotic regions, but  for $\bar{r}\sim 1$ its order is not 
far from unit
(though it is exponentially damped for large values of $\bar{r}$). 
Therefore, for a compact object with 
radius $R_s\approx 10$~km whose effective potential happens to 
possess 
a bound solution, 
it would take only a few milliseconds (according to far away 
observers) 
for the vacuum energy density to become {\it overwhelmingly 
dominant}
over classical matter densities as high as 
$10^{14-17}~{\rm g}/{\rm cm}^3$.
This suggests that the transition from classical to vacuum 
dominance can
be very sharp.
In spite of this, we emphasize that the {\it total}
vacuum energy is conserved (as it should in this static 
scenario where 
backreaction is not considered) due to the fact that the 
volume integral of 
$h(\bar r)$ vanishes. 


{\bf Final Comments:}
Although we have been talking about ``vacuum dominance'' of 
quantum fields,
we would like to clarify that 
the appearance of 
exponentially growing modes  already occurs in the 
classical level. This is evident from our argument that 
the
modes given in
Eq.~(\ref{voutmode}) are necessary in order to expand 
an arbitrary (classical) field configuration. Therefore, 
the very same situations which trigger the 
exponential growth of the vacuum energy 
density will {\it most likely} lead to an
exponential growth of the classical energy 
density stored in an {\it initially non-vanishing} 
field configuration. Here, however, we have opted to 
focus on the quantum version of the effect since it
is triggered even in the absence 
of any initial classical perturbation,
which makes its consequences 
more robust than the ones coming 
from its classical 
counterpart. 
The existence of initial classical perturbations can 
change the details but not the conclusion about whether 
or not the 
compact object will face 
the exponential instability, which is the main 
issue here.

The vacuum-awakening effect unveils a quite 
interesting interplay between semiclassical 
gravity and observational astrophysics.
Concerning field theorists, the observation of some 
relativistic stars
may be used to rule out the existence
of certain fields in Nature. For instance, according to our 
Fig.~\ref{Fig2PRL},
the observation of a {\it stable} and approximately spinless
cold neutron star with mass $M$ and radius $R_s$
(with an approximately parabolic density profile) rules out 
the existence of massless scalar fields with couplings $\xi$
for which $(\xi, M/R_s)$ lie in the dark-gray  region. 
Considering 
that $95\%$ of the energy content of the Universe is 
unknown, finding ways of testing the existence of 
{\it free} fields (for which there would be no 
reason to assume an 
initial classical perturbation) is very welcome.
Finally, concerning astrophysicists,
the awakening of the vacuum energy of certain fields may 
determine the ultimate fate of some relativistic stars. 
This will eventually depend on how the vacuum energy 
backreacts 
on the spacetime, as ruled by the
semiclassical Einstein equations.
The vacuum-driven evolution may lead, e.g., to (i) 
a complete collapse, hiding the negative-effective-potential 
region inside an event horizon, or (ii) an explosion, 
ejecting mass and bringing the effective potential back to a 
profile admitting no bound solution (or possibly 
a combination of both). Only a detailed numerical analysis, 
beyond the scope of this article, can decide what will 
happen once backreaction is taken into account.
Either way, we would have an unusual situation where the 
vacuum would be responsible for an event of astrophysical 
proportions.

\acknowledgments

The authors acknowledge partial financial support from 
Funda\c{c}\~ao de Amparo \`a Pesquisa do Estado de  S\~ao Paulo 
(FAPESP)
and Conselho Nacional de 
Desenvolvimento Cient\'\i fico e Tecnol\'ogico (CNPq). We also 
thank the anonymous referee for his/her valuable comments which 
helped to 
improve this version of the manuscript.

\end{document}